\documentclass{aastex6}

\usepackage{natbib}

\newcommand{\myemail}{ygong@pmo.ac.cn}

\newcommand{\kms}{km\,s$^{-1}$}

\begin{document}

%% LaTeX will automatically break titles if they run longer than
%% one line. However, you may use \\ to force a line break if
%% you desire.

\title{L1188: a promising candidate of cloud-cloud collision triggering the formation of the low- and intermediate-mass stars}

%% Use \author, \affil, plus the \and command to format author and affiliation 
%% information.  If done correctly the peer review system will be able to
%% automatically put the author and affiliation information from the manuscript
%% and save the corresponding author the trouble of entering it by hand.
%%
%% The \affil should be used to document primary affiliations and the
%% \altaffil should be used for secondary affiliations, titles, or email.

%% Authors with the same affiliation can be grouped in a single
%% \author and \affil call.
\author{Yan Gong\altaffilmark{1}, Min Fang\altaffilmark{1}, Ruiqing Mao\altaffilmark{1}, Shaobo Zhang\altaffilmark{1}, Yuan Wang\altaffilmark{1}, Yang Su\altaffilmark{1}, Xuepeng. Chen\altaffilmark{1}, Ji Yang\altaffilmark{1}, Hongchi Wang\altaffilmark{1}, Dengrong Lu\altaffilmark{1}}

%% Notice that each of these authors has alternate affiliations, which
%% are identified by the \altaffilmark after each name.  Specify alternate
%% affiliation information with \altaffiltext, with one command per each
%% affiliation.
\altaffiltext{1}{Purple Mountain Observatory \& Key Laboratory of Radio Astronomy, Chinese Academy of Sciences, 2 West Beijing Road, 210008 Nanjing, PR China, \myemail}

%% Mark off the abstract in the ``abstract'' environment. 
\begin{abstract}
We present a new large-scale (2\arcdeg$\times$2\arcdeg) simultaneous $^{12}$CO, $^{13}$CO, and C$^{18}$O ($J$=1$-$0) mapping of L1188 with the PMO 13.7-m telescope. Our observations have revealed that L1188 consists of two nearly orthogonal filamentary molecular clouds at two clearly separated velocities. Toward the intersection showing large velocity spreads, we find several bridging features connecting the two clouds in velocity, and an open arc structure which exhibits high excitation temperatures, enhanced $^{12}$CO and $^{13}$CO emission, and broad $^{12}$CO line wings. This agrees with the scenario that the two clouds are colliding with each other. The distribution of young stellar object (YSO) candidates implies an enhancement of star formation in the intersection of the two clouds. We suggest that a cloud-cloud collision happened in L1188 about 1~Myr ago, possibly triggering the formation of low- and intermediate-mass YSOs in the intersection.
\end{abstract}

%% Keywords should appear after the \end{abstract} command. 
%% See the online documentation for the full list of available subject
%% keywords and the rules for their use.

%% From the front matter, we move on to the body of the paper.
%% Sections are demarcated by \section and \subsection, respectively.
%% Observe the use of the LaTeX \label
%% command after the \subsection to give a symbolic KEY to the
%% subsection for cross-referencing in a \ref command.
%% You can use LaTeX's \ref and \label commands to keep track of
%% cross-references to sections, equations, tables, and figures.
%% That way, if you change the order of any elements, LaTeX will
%% automatically renumber them.

%% We recommend that authors also use the natbib \citep
%% and \citet commands to identify citations.  The citations are
%% tied to the reference list via symbolic KEYs. The KEY corresponds
%% to the KEY in the \bibitem in the reference list below. 

\section{Introduction} \label{sec:intro}
Triggered star formation sustains, amplifies, and disperses what large-scale instabilities begin \citep{1998ASPC..148..150E}. Cloud-cloud collisions are an important triggering mechanism, because such collisions are believed to be frequent in Milky-Way-like disk galaxies \citep{2009ApJ...700..358T} and are expected to induce starbursts in galaxies \citep{1990ApJ...349..480O}. As shown by hydrodynamics simulations \citep[e.g.,][]{1992PASJ...44..203H,2010MNRAS.405.1431A}, a cloud-cloud collision induces the formation of dense cores in the shock-compressed interface, and these dense cores are prone to gravitational instability leading to the formation of nascent stars. Recent magnetohydrodynamic simulations have shown that a cloud-cloud collision will lead to a large effective Jeans mass in the compressed layer, favoring the formation of massive cloud cores \citep{2013ApJ...774L..31I}. The analytical theory of \citet{2000ApJ...536..173T} argued if collisions are as frequent as several times per orbital period, cloud-cloud collisions could reproduce global star formation rates of galaxies and the Kennicutt-Schmidt relation \citep{1998ARA&A..36..189K}. However, good cloud-cloud collision candidates, including Westerlund 2 \citep{2009ApJ...696L.115F}, NGC 3603 \citep{2014ApJ...780...36F}, the Trifid Nebula M20 \citep{2011ApJ...738...46T}, RCW 120 \citep{2015ApJ...806....7T}, RCW 38 \citep{2016ApJ...820...26F}, and N159W-S in the Large Magellanic Cloud \citep{2015ApJ...807L...4F}, are still rare in observational aspects. In this letter, we present a new promising cloud-cloud collision candidate L1188 which is likely to trigger the formation of low- and intermediate-mass stars.
%the Serpens Main cluster \citep{2011A&A...528A..50D}

L1188, known as a dark cloud complex \citep{1962ApJS....7....1L}, is located at the periphery of the Cepheus Bubble \citep[see Fig.~21 of][]{2008hsf1.book..136K}. Previous observations have found several dense cores and young stellar objects (YSOs) in L1188 \citep{1995A&A...300..525A,2004BaltA..13..470K,2013AN....334..924M,2013AN....334..920V}, indicative of on-going star formation. On the other hand, there are rare CO studies toward L1188. An area of 74\arcmin$\times$44\arcmin\, was firstly studied in $^{13}$CO (1$-$0) with the Nagoya-4 m telescope \citep{1995A&A...300..525A}. The whole region was mapped by the FCRAO $^{12}$CO (1$-$0) outer galaxy survey \citep{1998ApJS..115..241H} and the Nagoya-4 m $^{13}$CO (1$-$0) survey \citep{1997ApJS..110...21Y}, but there is lack of published further analysis toward L1188. Here, we will present new multiple line observations toward this region in the $J$=1$-$0 transitions of the three main CO isotopologues ($^{12}$CO, $^{13}$CO, and C$^{18}$O), and detailed morphological and kinematic analysis.

%L1188 is also probably associated with the star-forming region NGC 7129 \citep{2001ApJ...560..841M} due to their similar Galactic longitudes and velocities. and 890$\pm$50 pc \citep{2013ApJ...769...15X} 

\section{Observations}
As part of the Milky Way Imaging Scroll Painting (MWISP\footnote{http://www.radioast.nsdc.cn/mwisp.php}) project, we carried out large-scale (2\arcdeg$\times$2\arcdeg) simultaneous observations of $^{12}$CO (1$-$0), $^{13}$CO (1$-$0), and C$^{18}$O (1$-$0) toward the L1188 molecular cloud complex with the Purple Mountain Observatory (PMO) 13.7-m telescope during 2011 November 13 to 2014 April 1. The $3\times$3-beam sideband separation Superconducting Spectroscopic Array Receiver \citep[SSAR,][]{pmo2} was used as front end, while a set of 18 fast Fourier transform spectrometers (FFTSs) were used as backend for signals from both sidebands. Each FFTS with a bandwidth of 1 GHz provides 16384 channels, resulting in a spectral resolution of 61 kHz, equivalent to a velocity resolution of $\sim$0.17~\kms\, at 110 GHz. Our observations cover a total of 16 cells, each of which is of size 30\arcmin$\times$30\arcmin\, and was observed with the On-The-Fly mode at a scanning rate of 50\arcsec\, per second and a dump time of 0.3 second. In order to reduce the scanning effects, each cell was mapped at least twice, along the Galactic longitude and the Galactic latitude, respectively. The observations encompasses a total of $\sim$72 observing hours.

The standard chopper-wheel method was used to calibrate the antenna temperature \citep{1976ApJS...30..247U}. We used the relationship $T_{\rm mb}=T_{\rm A} /B_{\rm eff}$ to convert the antenna temperature ($T_{\rm A}$) to the main beam temperature ($T_{\rm mb}$), where the main beam efficiency ($B_{\rm eff}$) is 44\% at 115 GHz and 48\% at 110 GHz according to the telescope status report\footnote{http://www.radioast.nsdc.cn/zhuangtaibaogao.php}. The calibration errors are estimated to be within 10\%. Typical system temperatures were 214---340 K at the upper sideband, and 126---176 K at the lower sideband. The resulting typical sensitivity is about 0.5 K ($T_{\rm mb}$) for $^{12}$CO (1$-$0) at a channel width of $\sim$0.16~\kms, and 0.3 K ($T_{\rm mb}$) for $^{13}$CO (1$-$0) and C$^{18}$O (1$-$0) at a channel width of $\sim$0.17~\kms. The half-power beam widths (HPBW) are about 55\arcsec\,and 52\arcsec\, at 110 GHz and 115 GHz, respectively. The pointing accuracy was accurate to $\sim$5\arcsec. Velocities are all given with respect to the local standard of rest (LSR) in this work. Data analysis was performed with the GILDAS data reduction package\footnote{http://www.iram.fr/IRAMFR/GILDAS} \citep[see e.g.][for details]{2016A&A...588A.104G}.

\section{Results}
\subsection{Distance}\label{sec.dis}
\begin{figure*}[!htbp]
\centering
\includegraphics[width = 0.6 \textwidth]{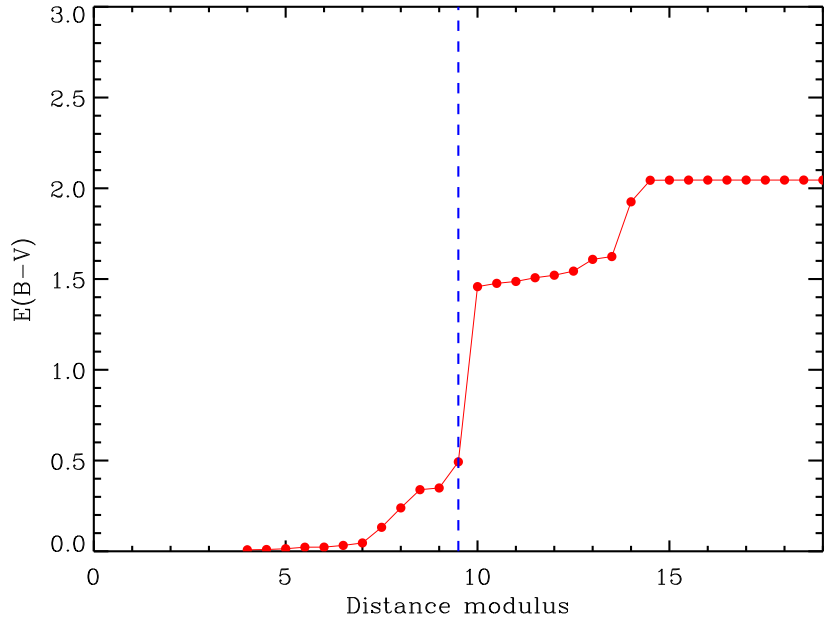}
\caption{{The median cumulative reddening in each distance modulus bin within the selected regions (see Sect.~\ref{sec.dis}). The dash line marks the distance that we use for L1188.} \label{Fig:ext}}
\end{figure*}

The distance is a fundamental parameter for deriving physical properties. However, there is no distance measurement toward L1188 so far. Here, we perform the distance estimate toward L1188 based on the 3D extinction map from \citet{2015ApJ...810...25G}. With 5-band $grizy$ Pan-STARRS~1 photometry and 3-band 2MASS $JHK_{\rm s}$ photometry of stars embedded in the dust, \citet{2015ApJ...810...25G} trace the extinction on 7\arcmin\,scales out to a distance of several kpc, by simultaneously inferring stellar distance, stellar type, and the reddening along the line of sight. We select two regions with high $^{13}$CO intensities integrated from $-$15 to $-$5~\kms. The regions are centered at the Galactic coordinates ($l$=106.0\arcdeg, $b$=3.9\arcdeg) and ($l$=105.9\arcdeg, $b$=4.2\arcdeg) with a radius of 0.15\arcdeg. In Fig.~\ref{Fig:ext}, we show the median cumulative reddening in each distance modulus (DM) bin within the selected regions. We note two rapid increases in this figure. One is at DM$\sim$9.5 ($\sim$800\,pc), and the other is at DM$\sim$14 (6500\,pc). L1188 has a velocity of around $-$10~\kms, in favor of the near distance. Thus, the rapid increase at the distance of $\sim$ 800\,pc should be due to the dust reddening in our studied molecular clouds. This also confirms the hypothesis, proposed by \citet{1995A&A...300..525A}, that L1188 is associated with L1204/S140, the distance of which was measured to be 764$\pm$27 pc \citep{2008PASJ...60..961H}. Hereafter, we take 800\,pc as the distance of L1188. 
%On basis of IRAS dust continuum emission morphology, \citet{1995A&A...300..525A} suggested that L1188 might be associated with L1204, the distance of which is found to be 764$\pm$27 pc \citep{2008PASJ...60..961H}. Our distance estimate confirms the argument. 

\subsection{Morphology and kinematics}\label{sec.mor}
\begin{figure*}[!htbp]
\centering
\includegraphics[width = 0.95 \textwidth]{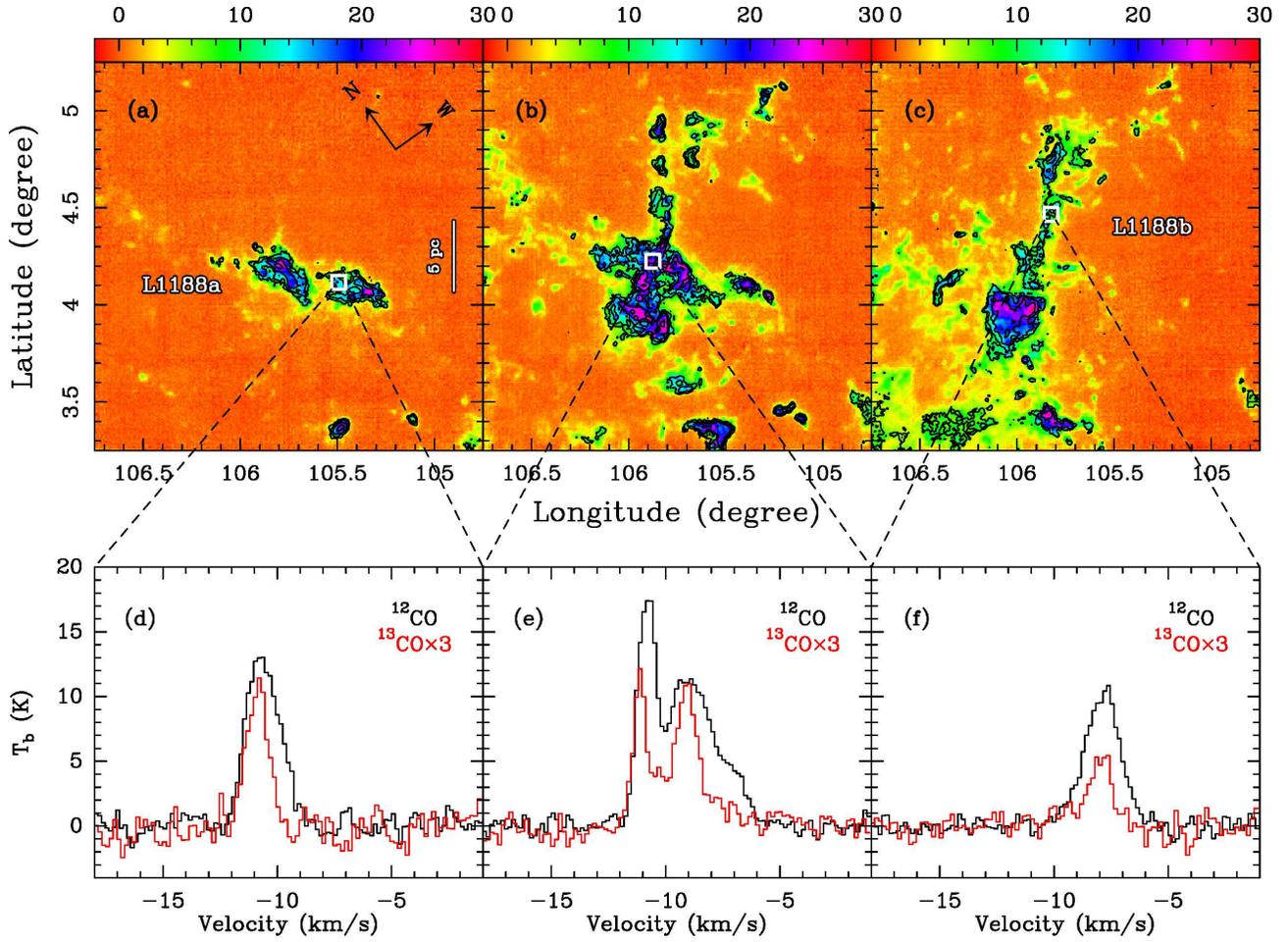}
\caption{{(a) $^{13}$CO integrated intensity contours overlaid on $^{12}$CO integrated intensity map. Both integrated velocity ranges are from $-$13.1 to $-$10.4~\kms. The lowest contour is 1 K~\kms\,(5$\sigma$), and each contour is twice the previous one. (b) The same as Fig.~\ref{Fig:mor}a but with the integrated velocity range from $-$10.4 to $-$8.0~\kms. (c) The same as Fig.~\ref{Fig:mor}a but with the integrated velocity range from $-$8.0 to $-$5.5~\kms. (d---f) The typical $^{12}$CO (black) and $^{13}$CO (red) spectra of the positions indicated by the white squares in Figs.~\ref{Fig:mor}a---\ref{Fig:mor}c. } \label{Fig:mor}}
\end{figure*}

Figures~\ref{Fig:mor}a---\ref{Fig:mor}c show the intensity maps of $^{12}$CO and $^{13}$CO with three different integrated velocity ranges, i.e., $-$13.1 to $-$10.4~\kms, $-$10.4 to $-$8.0~\kms, and $-$8.0 to $-$5.5~\kms. The integrated intensity maps display a molecular cloud elongated from northeast to southwest in Fig.~\ref{Fig:mor}a, and a cross-like morphology in Fig.~\ref{Fig:mor}b, and a molecular cloud elongated from northwest to southeast in Fig.~\ref{Fig:mor}c. The distributions suggest that the L1188 molecular cloud complex consists of two filamentary molecular clouds, named as L1188a (Fig.~\ref{Fig:mor}a) and L1188b (Fig.~\ref{Fig:mor}c), which are perpendicular to each other. L1188a has a length of $\sim$13.1 pc and a width of 0.7---2.8 pc (defined as the full width of the $^{13}$CO integrated intensity at the 3$\sigma$ level), while L1188b has a length of $\sim$15.9 pc and a width of 0.5---4.4 pc. Figures~\ref{Fig:mor}d---\ref{Fig:mor}f give the typical spectra in L1188a, L1188b, and the intersection of the two clouds. The spectra demonstrate that both L1188a and L1188b show single-peaked line profiles with systemic velocities of about $-$10.5 and $-$8~\kms, respectively, while the intersection is characterised by two velocity components. Assuming local thermodynamic equilibrium (LTE), the $^{13}$CO optical depths of the two peaks in Fig.~\ref{Fig:mor}e are less than 0.5.

By assuming that all molecular gas is in LTE, the total gas mass of L1188 is estimated with the $^{13}$CO intensities integrated from $-$15 to $-$5~\kms. Based on the excitation temperature map derived from the optically thick $^{12}$CO line and the correction factor $\frac{\tau}{1-{\rm exp}(-\tau)}$ \citep[$\tau$ is the optical depth of $^{13}$CO,][]{1999ApJ...517..209G} derived from the $^{13}$CO peak intensity, we can derive H$_{2}$ column densities with a constant [$^{12}$C/$^{13}$C] isotopic ratio of 77 \citep{1994ARA&A..32..191W} and a constant [H$_{2}$/$^{12}$CO] abundance ratio of 1.1$\times 10^{4}$ \citep{1982ApJ...262..590F}. By integrating over the area with the $^{13}$CO (1$-$0) emission higher than 3$\sigma$ and assuming a mean molecular weight per hydrogen molecule to be 2.8, the total mass is found to be 3.9$\times 10^{3}$~$M_{\odot}$ which is more than twice the previous result \citep[1.8$\times 10^{3}$~M$_{\odot}$,][]{1995A&A...300..525A}. This discrepancy can be understood since \citet{1995A&A...300..525A} only took dense regions into account and did not apply the correction factor. Assuming that the two clouds make the same contribution to the mass of the intersection, we arrive at a mass of 1.2$\times 10^{3}$~$M_{\odot}$ for L1188a and 2.7$\times 10^{3}$~$M_{\odot}$ for L1188b.

\begin{figure*}[!htbp]
\centering
\includegraphics[width = 0.95 \textwidth]{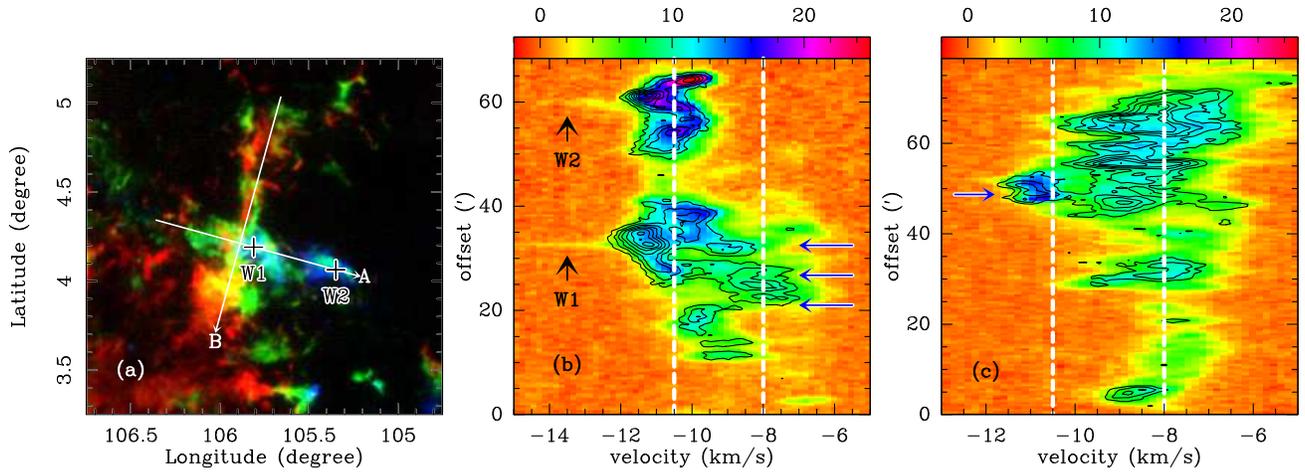}
\caption{{(a) A three-color integrated intensity $^{12}$CO (1$-$0) image of L1188 (red: integrated between $-$8.0 and $-$5.5~\kms; green: integrated between $-$10.4 and $-$8.0~\kms; blue: integrated between $-$13.1 and $-$10.4~\kms.). The two black crosses mark the positions (W1 and W2) showing $^{12}$CO high velocity wings. (b) The position-velocity diagram along the cut A indicated in Fig.~\ref{Fig:pv}a. $^{13}$CO black contours overlaid on the $^{12}$CO image. The color bar represents $^{12}$CO intensities in units of K. $^{13}$CO contours start at 0.9~K and increase by 0.9 K. The $^{12}$CO high velocity wings are pointed out by the black arrows. (c) The same as Fig.~\ref{Fig:pv}b but along the cut B indicated in Fig.~\ref{Fig:pv}a. The two white dashed lines represent the systemic velocity of L1188a and L1188b in Figs.~\ref{Fig:pv}b---\ref{Fig:pv}c. The bridging features are indicated by blue arrows in Figs.~\ref{Fig:pv}b and \ref{Fig:pv}c.} \label{Fig:pv}}
\end{figure*}

In order to investigate the velocity structure, we make position-velocity diagrams along the long axes of L1188a and L1188b shown in Fig.~\ref{Fig:pv}. In both $^{12}$CO and $^{13}$CO emission, one can clearly see bridging features in velocity between the two clouds. Although it is hard to separate such features individually from $^{12}$CO image only, $^{13}$CO contours can help to pick out at least four bridging features, indicated by blue arrows in Figs.~\ref{Fig:pv}b and \ref{Fig:pv}c. Such bridging features are a robust verification of the association of the two clouds \citep[e.g.,][]{2015MNRAS.454.1634H,2015MNRAS.450...10H,2016ApJ...820...26F}. Furthermore, we can see that both L1188a and L1188b have small velocity spreads of $\sim$2~\kms\, at the 3$\sigma$ level of $^{13}$CO emission. In contrast, the intersection of the two clouds shows a larger velocity spread of $\sim$5~\kms, indicating that molecular gas in this region is more turbulent. The region showing such a large velocity spread has a size of 20\arcmin, corresponding to a physical scale of 4.7 pc. 

\begin{figure*}[!htbp]
\centering
\includegraphics[width = 0.95 \textwidth]{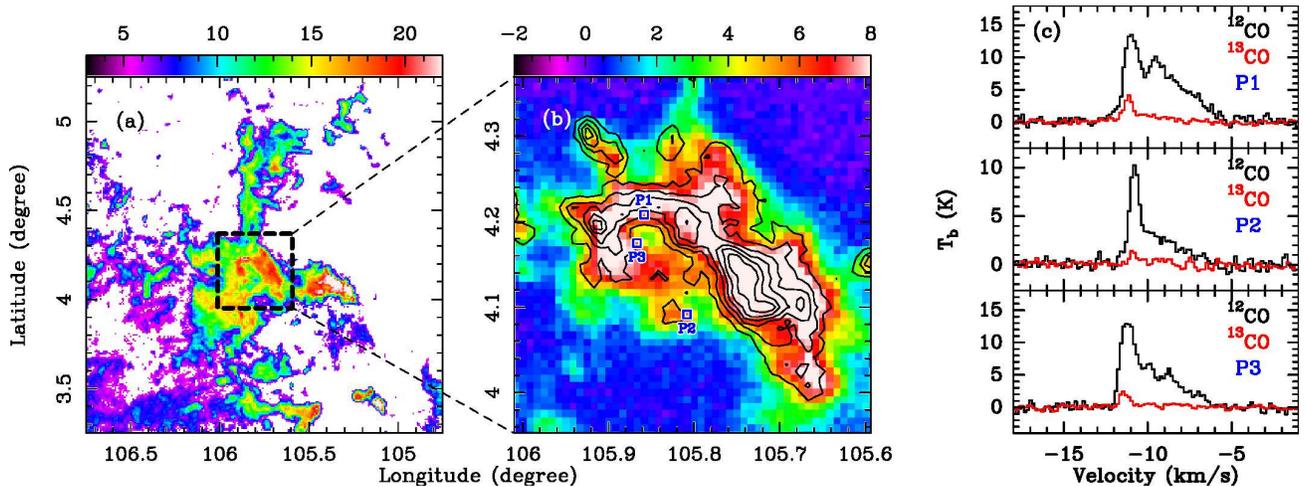}
\caption{{(a) The excitation temperature map derived from the peak intensity of $^{12}$CO (1$-$0), clipped at the 5$\sigma$ level. The color bar represents the excitation temperature in units of K. (b) Zoom in the region indicated by the black box in Fig.~\ref{Fig:coll}a. The $^{12}$CO integrated intensity map overlaid with the $^{13}$CO integrated intensity contours. Both integrated velocity ranges extend from $-$11.0 to $-$10.4 \kms. The color bar represents the $^{12}$CO integrated intensity in units of K~\kms. The contours start at 0.5~K~\kms\, and increase by 0.5~K~\kms. (c) The $^{12}$CO (black) and $^{13}$CO (red) spectra of the positions (P1, P2, and P3) indicated by the blue squares in Fig.~\ref{Fig:coll}b.} \label{Fig:coll}}
\end{figure*}

Figure~\ref{Fig:coll}a shows the $^{12}$CO excitation temperature map. In the intersection, we find an open arc structure which has excitation temperatures (18---22~K) higher than those ($<$15~K) of ambient gas. The radius of curvature of the arc structure is found to be about 0.6 pc. In addition, the arc structure appears to be enhanced in both $^{12}$CO and $^{13}$CO emission (see Fig.~\ref{Fig:coll}b). Toward the arc structure, we find widespread emission showing broad $^{12}$CO line wings (see three examples in Fig.~\ref{Fig:coll}c), indicating that molecular gas in the arc structure is perturbed by shocks.

\subsection{Star formation in the intersection of L1188a and L1188b}\label{sec.sf}
\begin{figure*}[!htbp]
\centering
\includegraphics[width = 0.95 \textwidth]{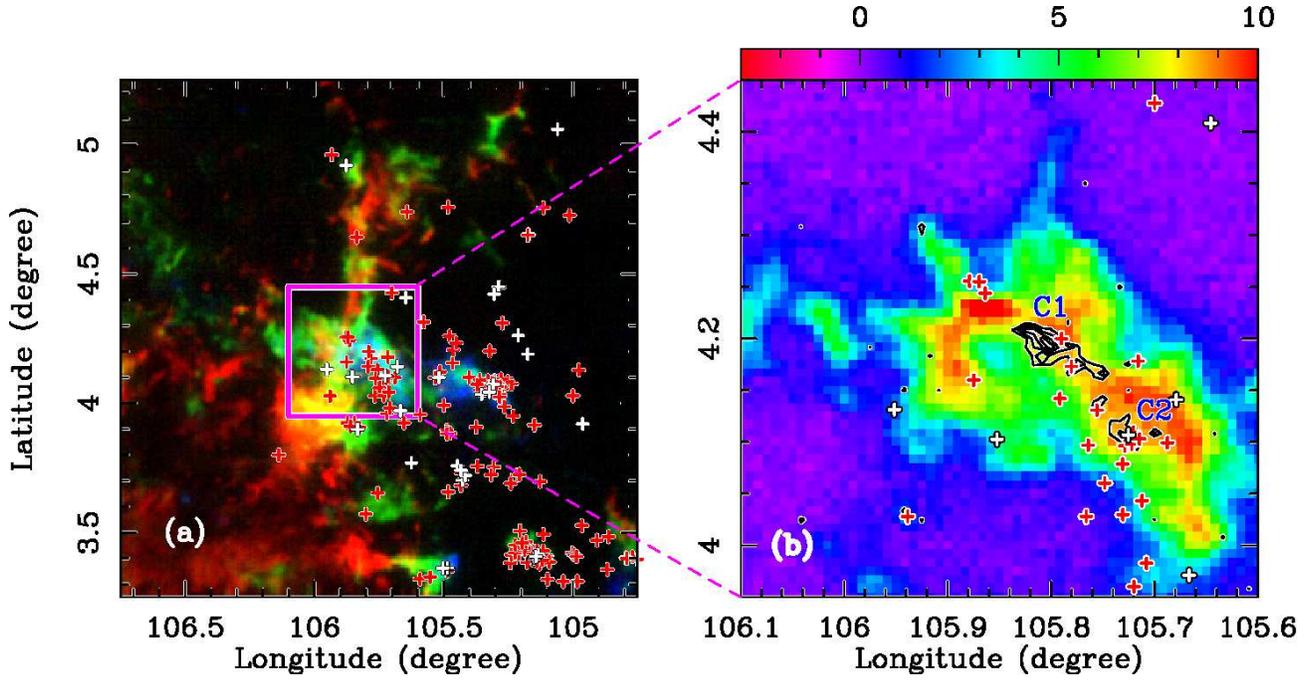}
\caption{{(a) The same as Fig.~\ref{Fig:pv}a but overlaid with the YSO distribution. (b) The $^{12}$CO integrated intensity map overlaid with the C$^{18}$O integrated intensity contours. $^{12}$CO is integrated from $-$11.0 to $-$10.4~\kms, while C$^{18}$O is integrated from $-$12 to $-$10~\kms. The color bar represents the $^{12}$CO integrated intensity in units of K~\kms. The contours start at 0.6~K~\kms\,(3$\sigma$) and increase by 0.4~K~\kms. Two C$^{18}$O cores are labeled as C1 and C2 in this panel. In both panels, Class I and Class II YSO candidates are marked with white and red crosses, respectively.} \label{Fig:sf}}
\end{figure*}

We use the WISE data \citep{2010AJ....140.1868W} and the criteria introduced by \citet{2012ApJ...744..130K} to search for class I and class II YSOs in the observed region. As a result, we found 228 YSO candidates, among which 31 are Class I and 197 are Class II. Figure~\ref{Fig:sf}a shows the distribution of these YSO candidates. In L1188, there are several regions showing overdense YSO candidates. What we are interested is the one located at the intersection of L1188a and L1188b. In this intersection, twenty-eight YSO candidates, including 3 class I and 25 class II YSO candidates, are found to be clustered around the arc structure. This is indicative of enhanced star formation. In the arc structure, we also discover two C$^{18}$O cores, labeled as C1 and C2 in Fig.~\ref{Fig:sf}b. Their deconvolved radii are found to be 0.22 pc and 0.16 pc. By assuming LTE and a constant [$^{16}$O/$^{18}$O] isotopic ratio of 560 \citep{1994ARA&A..32..191W}, their core masses are found to be 35~$M_{\odot}$ and 15~$M_{\odot}$ by summing up all column densities inside the contour of 0.6~K~\kms (see Fig.~\ref{Fig:sf}b). C1 is more elongated and massive than C2. Compared with the empirical mass-size relationship for the formation of massive stars \citep{2010ApJ...716..433K}, the two cores are unlikely to form massive stars. The two $^{12}$CO high velocity wings (W1 and W2) revealed in Fig.~\ref{Fig:pv}b are indicative of outflows driven by YSOs. From Fig.~\ref{Fig:pv}a, W1 is likely associated with C1. However, we cannot resolve the outflows with our data due to the limited angular resolution and the contamination from ambient clouds.

%In addition, we find that there is an infrared bow-shock nebula around the selected Class II YSO candidate J221744.99+615537.0 which also lies to the north of the arc structure (see Fig.~\ref{Fig:sf}c). The bow-shock nebula is likely due to the interaction between J221744.99+615537.0 and ambient molecular gas.

%These results indicates that the two clouds are colliding with each other.
%The region with this feature is found to have a size of about 4.7~pc (20\arcmin) in Fig.~\ref{Fig:pv}b and \ref{Fig:pv}c. 
%More details of outflows and cores will be published separately.
%Based on the NVSS survey \citep{1998AJ....115.1693C}, We find a continuum point source ($l$=105.734\arcdeg, $b$=4.176\arcdeg) in the intersection. The ionizing source has a flux density of 18.6~mJy at 1.4 GHz. Assuming that the continuum emission is optically thin, we obtain the ionizing photon rate of 1.0$\times 10^{45}$~s$^{-1}$ with an typical electron temperature of 8000 K. Comparing the value with Table II of \citet{1973AJ.....78..929P}, we suggest that the source is a weak H{\scriptsize II} region ionized by a B2 star with a mass of $\sim$10~M$\odot$.
\section{Discussion}
In Sect.~\ref{sec.mor}, we suggest that L1188 consists of two nearly orthogonal filamentary molecular clouds. Although the morphology can be alternatively explained by cores accretion from the surrounding filament material which was believed to play an important role in the formation of the embedded Serpens south protocluster \citep{2013ApJ...766..115K}, this doesn't seem to be the case for L1188. Such filamentary accretion flows will display self-absorption in optically thick emission and a single peaked profile in optically thin emission, which is not found in our data. Instead, our $^{13}$CO data has low optical depths ($<$0.5) and show two distinct peaks, which supports that the two velocity components arise from two clouds rather than accretion flows. We also find several pieces of evidence to support the idea that L1188a and L1188b are colliding with each other. Firstly, bridging features connect the two clouds in velocity, which points out that the two clouds are interacting \citep{2015MNRAS.454.1634H,2015MNRAS.450...10H}. Secondly, the intersection of the two clouds show large velocity spreads, indicative of more turbulent motions. This agrees with results given by simulations of cloud-cloud collisions \citep[e.g.,][]{2013ApJ...774L..31I}. Thirdly, an arc structure exhibiting high excitation temperatures, enhanced $^{12}$CO and $^{13}$CO emission, and broad $^{12}$CO line wings is found in the intersection of the two clouds. Such a structure could result from stellar feedback. However, we do not find H{\scriptsize II} regions or YSOs toward the center of the arc structure to our knowledge. Instead, the open arc structure has been successfully created by simulations of cloud-cloud collisions where the arc-like layer is formed due to the shock compression \citep[e.g.,][]{1992PASJ...44..203H,2014ApJ...792...63T}. Toward the arc structure, the widespread gas showing broad $^{12}$CO line wings reinforce the presence of shocks. Furthermore, there is lack of YSOs around the positions showing such wings, indicating that the shocks arise from the cloud-cloud collision. We note that such wing profile is seldom reported in the other cloud-cloud candidates mentioned in Sect.~\ref{sec:intro}.

From the systemic velocities of the two clouds, their velocity difference is about 2.5~\kms\,which is a lower limit of the relative collision velocity due to the projection effect. Previous studies suggest that the velocity spread in the shocked layer is similar to the relative collision velocity \citep{2013ApJ...774L..31I,2015ApJ...807L...4F}. We thus roughly assume the relative collision velocity to be 5~\kms. The collision velocity is also in consistent with simulations of small clouds which predicted an open arc dense structure at the assumed relative collision velocity \citep{2014ApJ...792...63T}. Compared with the velocity difference, the relative collision velocity implies that the relative motion of the two clouds is about 60\arcdeg\,to the line of sight. Assuming an effective radius of 5 pc for L1188, we estimate that a mass of 2.4$\times 10^{4}$~$M_{\odot}$ is required to gravitationally bind the two clouds, which is six times the observed LTE mass (3.9$\times 10^{3}$~$M_{\odot}$). Thus, the total mass of L1188 is too small to gravitationally bind the two clouds.

Given that star forming activities are found in the intersection, we propose a scenario that star formation is triggered by a collision of molecular clouds in L1188. The relative collision velocity of 5~\kms\,gives a transverse velocity of 4.3~\kms. Assuming that the perturbed region showing large velocity spreads in Fig.~\ref{Fig:pv}b is ascribable to the collision, this implies that L1188a and L1188b collided with each other about a timescale of 4.7 pc divided by 4.3~\kms, i.e., $sim$1 Myr ago. Such a timescale is longer than the lifetime ($\sim$0.5 Myr) of Class I YSOs and comparable to the lifetime (2$\pm$1 Myr) of Class II YSOs \citep{2009ApJS..181..321E}, which means that the YSO candidates in the region are possible to form after the collision between L1188a and L1188b. This implies that these stars are possible to be triggered by the collision. Together with the geometric simplicity of L1188, the cloud complex is unlikely to form massive stars, which makes it one of the best candidates to further study cloud-cloud collision triggering the formation of the low- and intermediate-mass stars. 

%If we consider a typical low-mass star ($\sim$2~M$_{\odot}$) to be triggered at the beginning of the collision, this gives an average mass accretion rate of $\sim$1$\times 10^{-6}$ M$_{\odot}$~yr$^{-1}$ which is higher than the typical mass accretion rate of Class II YSOs \citep[$\lesssim 10^{-7}$~M$_{\odot}$~yr$^{-1}$,][]{1998ApJ...492..323G} by almost an order of magnitude. This indicates that the triggering mechanism through cloud-cloud collisions may increase the mass accretion rate in star-forming processes. 

%Compared with the aforementioned candidates (see Sect.~\ref{sec:intro}), L1188 is the only source showing the open arc dense structure predicated by \citet{2014ApJ...792...63T}. In combination with the unique cross-like morphology, L1188 is thus an exceptional target of triggered low-mass star formation by cloud-cloud collisions.

%Assuming that the perturbed regions are due to shocks induced by a cloud-cloud collision, we estimate the shock propagation velocity to be about 4.7~pc/3.7~Myr$\sim$1.2~\kms. 

\section{Summary and conclusions}
Based on the 3D extinction map, we perform the distance estimate toward L1188 which is found to be $\sim$800 pc. We present a new large-scale (2\arcdeg$\times$2\arcdeg) mapping of L1188 in the three CO isotopic lines with the PMO 13.7-m telescope. We unveil that L1188, displaying a cross-like morphology, consists of two nearly orthogonal filamentary molecular clouds (L1188a and L1188b) with systemic velocities of about $-$10.5 and $-$8.0~\kms. The total mass of L1188 is 3.9$\times 10^{3}$~M$_{\odot}$, unable to bind the two clouds. Toward the intersection of the two clouds which shows large velocity spreads, we find several bridging features connecting the two clouds in velocity, and an open arc structure which exhibits high excitation temperatures, enhanced $^{12}$CO and $^{13}$CO emission, and broad $^{12}$CO line wings. This is in consistent with the scenario that the two clouds are colliding with each other. An overdense distribution of young stellar object candidates is found in the intersection of the two clouds, indicative of enhanced star formation. We suggest that a cloud-cloud collision happened in L1188 about 1 Myr ago, possibly triggering the formation of low- and intermediate-mass stars in the intersection.

%% If you wish to include an acknowledgments section in your paper,
%% separate it off from the body of the text using the \acknowledgments
%% command.
\acknowledgments
We thank the referee for useful comments that improved the letter. We appreciate the assistance of the PMO 13.7-m operators during the observations. We acknowledge support by the MWISP team, the National Natural Science Foundation of China (NSFC) (grants nos. 11127903, 11233007, 10973040), and the Strategic Priority Research Program of the Chinese Academy of Sciences (grant no. XDB09000000). M.F., Y.W., and X.-P.C. were supported by the NSFC under grants 11203081, 11303097, and 11473069, respectively. This research made use of NASA's Astrophysics Data System.

%% To help institutions obtain information on the effectiveness of their 
%% telescopes the AAS Journals has created a group of keywords for telescope 
%% facilities. 

%% Following the acknowledgments section, use the following syntax and the
%% \facility{} macro to list the keywords of facilities used in the research 
%% for the paper.  Each keyword is check against the master list during
%% copy editing.  Individual instruments can be provided in parentheses,
%% after the keyword, but they are not verified.
\facilities{PMO 13.7-m}

%{\it Facilities:} \facility{Effelsberg-100 m}, \facility{ATCA}, \facility{JVLA}.

%% Appendix material should be preceded with a single \appendix command.
%% There should be a \section command for each appendix. Mark appendix
%% subsections with the same markup you use in the main body of the paper.

%% Each Appendix (indicated with \section) will be lettered A, B, C, etc.
%% The equation counter will reset when it encounters the \appendix
%% command and will number appendix equations (A1), (A2), etc.

\clearpage

%(c) A WISE three-color composite image (red: 22 $\mu$m; green: 12 $\mu$m; blue: 4.6 $\mu$m) around J221744.99+615537.0. In all panels, Class I and Class II YSO candidates are marked by the white and red crosses, respectively.

\bibliographystyle{apj}
%\bibliography{references}

\begin{thebibliography}{38}
\expandafter\ifx\csname natexlab\endcsname\relax\def\natexlab#1{#1}\fi

\bibitem[{{Abraham} {et~al.}(1995){Abraham}, {Dobashi}, {Mizuno}, \&
  {Fukui}}]{1995A&A...300..525A}
{Abraham}, P., {Dobashi}, K., {Mizuno}, A., \& {Fukui}, Y. 1995, \aap, 300, 525

\bibitem[{{Anathpindika}(2010)}]{2010MNRAS.405.1431A}
{Anathpindika}, S.~V. 2010, \mnras, 405, 1431

\bibitem[{{Elmegreen}(1998)}]{1998ASPC..148..150E}
{Elmegreen}, B.~G. 1998, in Astronomical Society of the Pacific Conference
  Series, Vol. 148, Origins, ed. C.~E. {Woodward}, J.~M. {Shull}, \& H.~A.
  {Thronson}, Jr., 150

\bibitem[{{Evans} {et~al.}(2009){Evans}, {Dunham}, {J{\o}rgensen}, {Enoch},
  {Mer{\'{\i}}n}, {van Dishoeck}, {Alcal{\'a}}, {Myers}, {Stapelfeldt},
  {Huard}, {Allen}, {Harvey}, {van Kempen}, {Blake}, {Koerner}, {Mundy},
  {Padgett}, \& {Sargent}}]{2009ApJS..181..321E}
{Evans}, II, N.~J., {et~al.} 2009, \apjs, 181, 321

\bibitem[{{Frerking} {et~al.}(1982){Frerking}, {Langer}, \&
  {Wilson}}]{1982ApJ...262..590F}
{Frerking}, M.~A., {Langer}, W.~D., \& {Wilson}, R.~W. 1982, \apj, 262, 590

\bibitem[{{Fukui} {et~al.}(2014){Fukui}, {Ohama}, {Hanaoka}, {Furukawa},
  {Torii}, {Dawson}, {Mizuno}, {Hasegawa}, {Fukuda}, {Soga}, {Moribe},
  {Kuroda}, {Hayakawa}, {Kawamura}, {Kuwahara}, {Yamamoto}, {Okuda}, {Onishi},
  {Maezawa}, \& {Mizuno}}]{2014ApJ...780...36F}
{Fukui}, Y., {et~al.} 2014, \apj, 780, 36

\bibitem[{{Fukui} {et~al.}(2015){Fukui}, {Harada}, {Tokuda}, {Morioka},
  {Onishi}, {Torii}, {Ohama}, {Hattori}, {Nayak}, {Meixner}, {Sewi{\l}o},
  {Indebetouw}, {Kawamura}, {Saigo}, {Yamamoto}, {Tachihara}, {Minamidani},
  {Inoue}, {Madden}, {Galametz}, {Lebouteiller}, {Mizuno}, \&
  {Chen}}]{2015ApJ...807L...4F}
---. 2015, \apjl, 807, L4

\bibitem[{{Fukui} {et~al.}(2016){Fukui}, {Torii}, {Ohama}, {Hasegawa},
  {Hattori}, {Sano}, {Ohashi}, {Fujii}, {Kuwahara}, {Mizuno}, {Dawson},
  {Yamamoto}, {Tachihara}, {Okuda}, {Onishi}, \&
  {Mizuno}}]{2016ApJ...820...26F}
---. 2016, \apj, 820, 26

\bibitem[{{Furukawa} {et~al.}(2009){Furukawa}, {Dawson}, {Ohama}, {Kawamura},
  {Mizuno}, {Onishi}, \& {Fukui}}]{2009ApJ...696L.115F}
{Furukawa}, N., {Dawson}, J.~R., {Ohama}, A., {Kawamura}, A., {Mizuno}, N.,
  {Onishi}, T., \& {Fukui}, Y. 2009, \apjl, 696, L115

\bibitem[{{Goldsmith} \& {Langer}(1999)}]{1999ApJ...517..209G}
{Goldsmith}, P.~F., \& {Langer}, W.~D. 1999, \apj, 517, 209

\bibitem[{{Gong} {et~al.}(2016){Gong}, {Mao}, {Fang}, {Zhang}, {Su}, {Yang},
  {Jiang}, {Xu}, {Wang}, {Wang}, {Lu}, \& {Sun}}]{2016A&A...588A.104G}
{Gong}, Y., {et~al.} 2016, \aap, 588, A104

\bibitem[{{Green} {et~al.}(2015){Green}, {Schlafly}, {Finkbeiner}, {Rix},
  {Martin}, {Burgett}, {Draper}, {Flewelling}, {Hodapp}, {Kaiser}, {Kudritzki},
  {Magnier}, {Metcalfe}, {Price}, {Tonry}, \&
  {Wainscoat}}]{2015ApJ...810...25G}
{Green}, G.~M., {et~al.} 2015, \apj, 810, 25

\bibitem[{{Habe} \& {Ohta}(1992)}]{1992PASJ...44..203H}
{Habe}, A., \& {Ohta}, K. 1992, \pasj, 44, 203

\bibitem[{{Haworth} {et~al.}(2015{\natexlab{a}}){Haworth}, {Shima}, {Tasker},
  {Fukui}, {Torii}, {Dale}, {Takahira}, \& {Habe}}]{2015MNRAS.454.1634H}
{Haworth}, T.~J., {Shima}, K., {Tasker}, E.~J., {Fukui}, Y., {Torii}, K.,
  {Dale}, J.~E., {Takahira}, K., \& {Habe}, A. 2015{\natexlab{a}}, \mnras, 454,
  1634

\bibitem[{{Haworth} {et~al.}(2015{\natexlab{b}}){Haworth}, {Tasker}, {Fukui},
  {Torii}, {Dale}, {Shima}, {Takahira}, {Habe}, \&
  {Hasegawa}}]{2015MNRAS.450...10H}
{Haworth}, T.~J., {et~al.} 2015{\natexlab{b}}, \mnras, 450, 10

\bibitem[{{Heyer} {et~al.}(1998){Heyer}, {Brunt}, {Snell}, {Howe}, {Schloerb},
  \& {Carpenter}}]{1998ApJS..115..241H}
{Heyer}, M.~H., {Brunt}, C., {Snell}, R.~L., {Howe}, J.~E., {Schloerb}, F.~P.,
  \& {Carpenter}, J.~M. 1998, \apjs, 115, 241

\bibitem[{{Hirota} {et~al.}(2008){Hirota}, {Ando}, {Bushimata}, {Choi},
  {Honma}, {Imai}, {Iwadate}, {Jike}, {Kameno}, {Kameya}, {Kamohara}, {Kan-Ya},
  {Kawaguchi}, {Kijima}, {Kim}, {Kobayashi}, {Kuji}, {Kurayama}, {Manabe},
  {Matsui}, {Matsumoto}, {Miyaji}, {Miyazaki}, {Nagayama}, {Nakagawa},
  {Namikawa}, {Nyu}, {Oh}, {Omodaka}, {Oyama}, {Sakai}, {Sasao}, {Sato},
  {Sato}, {Shibata}, {Tamura}, {Ueda}, \& {Yamashita}}]{2008PASJ...60..961H}
{Hirota}, T., {et~al.} 2008, \pasj, 60, 961

\bibitem[{{Inoue} \& {Fukui}(2013)}]{2013ApJ...774L..31I}
{Inoue}, T., \& {Fukui}, Y. 2013, \apjl, 774, L31

\bibitem[{{Kauffmann} {et~al.}(2010){Kauffmann}, {Pillai}, {Shetty}, {Myers},
  \& {Goodman}}]{2010ApJ...716..433K}
{Kauffmann}, J., {Pillai}, T., {Shetty}, R., {Myers}, P.~C., \& {Goodman},
  A.~A. 2010, \apj, 716, 433

\bibitem[{{Kennicutt}(1998)}]{1998ARA&A..36..189K}
{Kennicutt}, Jr., R.~C. 1998, \araa, 36, 189

\bibitem[{{Kirk} {et~al.}(2013){Kirk}, {Myers}, {Bourke}, {Gutermuth},
  {Hedden}, \& {Wilson}}]{2013ApJ...766..115K}
{Kirk}, H., {Myers}, P.~C., {Bourke}, T.~L., {Gutermuth}, R.~A., {Hedden}, A.,
  \& {Wilson}, G.~W. 2013, \apj, 766, 115

\bibitem[{{Koenig} {et~al.}(2012){Koenig}, {Leisawitz}, {Benford}, {Rebull},
  {Padgett}, \& {Assef}}]{2012ApJ...744..130K}
{Koenig}, X.~P., {Leisawitz}, D.~T., {Benford}, D.~J., {Rebull}, L.~M.,
  {Padgett}, D.~L., \& {Assef}, R.~J. 2012, \apj, 744, 130

\bibitem[{{K{\"o}nyves} {et~al.}(2004){K{\"o}nyves}, {Mo{\'o}r}, {Kiss}, \&
  {{\'A}brah{\'a}m}}]{2004BaltA..13..470K}
{K{\"o}nyves}, V., {Mo{\'o}r}, A., {Kiss}, C., \& {{\'A}brah{\'a}m}, P. 2004,
  Baltic Astronomy, 13, 470

\bibitem[{{Kun} {et~al.}(2008){Kun}, {Kiss}, \& {Balog}}]{2008hsf1.book..136K}
{Kun}, M., {Kiss}, Z.~T., \& {Balog}, Z. 2008, {Star Forming Regions in
  Cepheus}, ed. B.~{Reipurth}, 136

\bibitem[{{Lynds}(1962)}]{1962ApJS....7....1L}
{Lynds}, B.~T. 1962, \apjs, 7, 1

\bibitem[{{Marton } {et~al.}(2013){Marton }, {Vereb{\'e}lyi}, {Kiss}, \&
  {Smidla}}]{2013AN....334..924M}
{Marton }, G., {Vereb{\'e}lyi}, E., {Kiss}, C., \& {Smidla}, J. 2013,
  Astronomische Nachrichten, 334, 924

\bibitem[{{Olson} \& {Kwan}(1990)}]{1990ApJ...349..480O}
{Olson}, K.~M., \& {Kwan}, J. 1990, \apj, 349, 480

\bibitem[{{Shan} {et~al.}(2012){Shan}, {Yang}, {Shi}, {Yao}, {Zuo}, {Lin},
  {Chen}, {Zhang}, {Duan}, {Cao}, {Li}, {Li}, {Liu}, \& {Zhong}}]{pmo2}
{Shan}, W.~L., {et~al.} 2012, IEEE Transactions on Terahertz Science and
  Technology, 2, 593

\bibitem[{{Takahira} {et~al.}(2014){Takahira}, {Tasker}, \&
  {Habe}}]{2014ApJ...792...63T}
{Takahira}, K., {Tasker}, E.~J., \& {Habe}, A. 2014, \apj, 792, 63

\bibitem[{{Tan}(2000)}]{2000ApJ...536..173T}
{Tan}, J.~C. 2000, \apj, 536, 173

\bibitem[{{Tasker} \& {Tan}(2009)}]{2009ApJ...700..358T}
{Tasker}, E.~J., \& {Tan}, J.~C. 2009, \apj, 700, 358

\bibitem[{{Torii} {et~al.}(2011){Torii}, {Enokiya}, {Sano}, {Yoshiike},
  {Hanaoka}, {Ohama}, {Furukawa}, {Dawson}, {Moribe}, {Oishi}, {Nakashima},
  {Okuda}, {Yamamoto}, {Kawamura}, {Mizuno}, {Maezawa}, {Onishi}, {Mizuno}, \&
  {Fukui}}]{2011ApJ...738...46T}
{Torii}, K., {et~al.} 2011, \apj, 738, 46

\bibitem[{{Torii} {et~al.}(2015){Torii}, {Hasegawa}, {Hattori}, {Sano},
  {Ohama}, {Yamamoto}, {Tachihara}, {Soga}, {Shimizu}, {Okuda}, {Mizuno},
  {Onishi}, {Mizuno}, \& {Fukui}}]{2015ApJ...806....7T}
---. 2015, \apj, 806, 7

\bibitem[{{Ulich} \& {Haas}(1976)}]{1976ApJS...30..247U}
{Ulich}, B.~L., \& {Haas}, R.~W. 1976, \apjs, 30, 247

\bibitem[{{Vereb{\'e}lyi } {et~al.}(2013){Vereb{\'e}lyi }, {K{\"o}nyves},
  {Nikoli{\'c}}, {Kiss}, {Mo{\'o}r}, {{\'A}brah{\'a}m}, \&
  {Kun}}]{2013AN....334..920V}
{Vereb{\'e}lyi }, E., {K{\"o}nyves}, V., {Nikoli{\'c}}, S., {Kiss}, C.,
  {Mo{\'o}r}, A., {{\'A}brah{\'a}m}, P., \& {Kun}, M. 2013, Astronomische
  Nachrichten, 334, 920

\bibitem[{{Wilson} \& {Rood}(1994)}]{1994ARA&A..32..191W}
{Wilson}, T.~L., \& {Rood}, R. 1994, \araa, 32, 191

\bibitem[{{Wright} {et~al.}(2010){Wright}, {Eisenhardt}, {Mainzer}, {Ressler},
  {Cutri}, {Jarrett}, {Kirkpatrick}, {Padgett}, {McMillan}, {Skrutskie},
  {Stanford}, {Cohen}, {Walker}, {Mather}, {Leisawitz}, {Gautier}, {McLean},
  {Benford}, {Lonsdale}, {Blain}, {Mendez}, {Irace}, {Duval}, {Liu}, {Royer},
  {Heinrichsen}, {Howard}, {Shannon}, {Kendall}, {Walsh}, {Larsen}, {Cardon},
  {Schick}, {Schwalm}, {Abid}, {Fabinsky}, {Naes}, \&
  {Tsai}}]{2010AJ....140.1868W}
{Wright}, E.~L., {et~al.} 2010, \aj, 140, 1868

\bibitem[{{Yonekura} {et~al.}(1997){Yonekura}, {Dobashi}, {Mizuno}, {Ogawa}, \&
  {Fukui}}]{1997ApJS..110...21Y}
{Yonekura}, Y., {Dobashi}, K., {Mizuno}, A., {Ogawa}, H., \& {Fukui}, Y. 1997,
  \apjs, 110, 21

\end{thebibliography}

%\clearpage
%\appendix
%\section{Convert the flux density to the brightness temperature}
%
%
%

%% This command is needed to show the entire author+affilation list when
%% the collaboration and author truncation commands are used.  It has to
%% go at the end of the manuscript.
%\allauthors

%% Include this line if you are using the \added, \replaced, \deleted
%% commands to see a summary list of all changes at the end of the article.
%\listofchanges

\end{document}